\documentclass[aps,prd,twocolumn,showpacs,amsmath,amssymb,nofootinbib,eqsecnum, preprintnumbers]{revtex4-1}

\newcommand{\eps}{\varepsilon}

 \newcommand{\dm}{n}

\newcommand{\be}{\begin{equation}}
\newcommand{\ee}{\end{equation}}
\newcommand{\ba}{\begin{eqnarray}}
\newcommand{\ea}{\end{eqnarray}}

\newcommand{\beq}{\begin{equation}}
\newcommand{\eeq}{\end{equation}}
\newcommand{\beqa}{\begin{eqnarray}}
\newcommand{\eeqa}{\end{eqnarray}}
\newcommand{\nn}{\nonumber}

\newcommand{\cv}[1]{{\partial}_{#1}}

\newcommand{\ef}{E}

\newcommand{\KV}[1]{\xi_{(#1)}}
\newcommand{\KT}[1]{K_{(#1)}}

\newcommand{\A}[1]{A^{\!(#1)}}

\begin{document}
\title{On integrability of spinning particle motion in higher-dimensional rotating black hole spacetimes}

\author{David Kubiz\v n\'ak}
\email{dkubiznak@perimeterinstitute.ca}
\affiliation{DAMTP, University of Cambridge, Wilberforce Road, Cambridge CB3 0WA, UK}
\affiliation{Perimeter Institute, 31 Caroline St. N. Waterloo
Ontario, N2L 2Y5, Canada}

\author{Marco Cariglia}
\email{marco@iceb.ufop.br}
\affiliation{Universidade Federal de Ouro Preto, ICEB, Departamento de F\'isica.
  Campus Morro do Cruzeiro, Morro do Cruzeiro, 35400-000 - Ouro Preto, MG - Brasil}

\date{December 1, 2011}  

\begin{abstract}
We study the motion of a classical spinning particle (with spin degrees of freedom described by a vector of Grassmann variables) in higher-dimensional 
general rotating black hole spacetimes with a cosmological constant. In all dimensions $n$  we exhibit $n$ bosonic functionally independent 
integrals of spinning particle motion, corresponding to explicit and hidden symmetries generated from the principal conformal Killing--Yano tensor. 
Moreover, we demonstrate that in 4-, 5-, 6-, and 7-dimensional black hole spacetimes such integrals are in involution, proving the bosonic part of the motion integrable. We conjecture that the same conclusion remains valid in all higher dimensions.
Our result generalizes the 
result of Page et. al. [hep-th/0611083] on complete integrability of geodesic motion in these spacetimes.
\end{abstract}

\pacs{04.50.-h, 04.50.Gh, 04.70.Bw, 04.20.Jb}
\preprint{DAMTP-2011-80}

\maketitle

\section{Introduction}
{Complete integrability} and chaotic motion are two opposite characteristic behaviors of any dynamical system. Whereas  
integrability is exceptional, only a very few systems render this property, chaotic motion is generic. 
In the special case of Hamiltonian systems one has a notion of {\em complete Liouville integrability}, see, e.g., \cite{Arnold:book}.
This means that the equations of motion are `integrable by quadratures', i.e., that the solution can
be found after a finite number of steps involving algebraic operations and integration. Mathematically this translates to the 
requirement that the Hamiltonian system with $n$ degrees of freedom admits $n$ {\em functionally independent integrals of motion which are in involution}, i.e, their Poisson brackets mutually vanish.
This is possible only if the system admits a sufficient number of corresponding symmetries. 
Well known examples of completely integrable mechanical systems are, for example, the motion in the central potential, Neumann's oscillator, or the geodesic motion on an ellipsoid. In black hole physics one of the rather nontrivial examples of complete integrability is the integrability of geodesic motion in rotating black hole spacetimes 
in four \cite{Carter:1968pr} and higher \cite{PageEtal:2007, FrolovEtal:2007} dimensions. The symmetry responsible for this result is encoded in the hidden symmetry of the so called {\em principal conformal Killing--Yano} tensor \cite{FrolovKubiznak:2007, KubiznakFrolov:2007, KrtousEtal:2007jhep}.   

In this paper we study integrability of a new system.
Namely, we extend the result on complete integrability of geodesic motion in spherical higher-dimensional rotating black hole spacetimes \cite{ChenEtal:2006cqg} by including into consideration the 
particle's spin. We demonstrate complete integrability of the ``bosonic sector
of motion'' of a spinning particle described by the {\em supersymmetric
classical theory} \cite{BerezinMarinov:1977,Casalbuoni:1976,BarducciEtal:1976,BrinkEtal:1976,%
BrinkEtal:1977,RietdijkHolten:1989,GibbonsEtal:1993,Tanimoto:1995,AhmedovAliev:2009,NgomeEtal:2010} where the spin degrees of freedom are described by Grassmann (anticommuting) variables.\footnote{In this paper we do not discuss fermionic equations of motion. Such equations are already of the first order, however their integrability might require a separate analysis.}
Such a theory is physically very interesting. On one side it provides a semi-classical description of Dirac's theory of spin $\frac{1}{2}$ fermions,  while on the other side it is closely related to the classical general relativistic description of extended objects with spin, described by Papapetrou's equations \cite{Papapetrou:1951}.
Our work thus provides a new example of integrability which is both nontrivial and physically interesting.

In the next two sections we briefly recapitulate the classical theory of spinning particles and review the basic properties of higher-dimensional rotating black hole spacetimes. 
Then we present the new bosonic (quadratic in momenta) integrals of spinning particle motion in all dimensions and demonstrate that in 4-, 5-, 6-, and 7-dimensional black hole spacetimes such integrals guarantee complete integrability of the bosonic sector of the spinning particle motion. 
Our conjecture is that the same
conclusion remains true in any dimension.
We conclude with open questions and future possible directions. The detailed calculations will be presented elsewhere \cite{CarigliaKubiznak:2011}.

\section{Classical spinning particle}
We describe the spinning particle by a classical theory 
which is a worldline supersymmetric extension of the ordinary relativistic point-particle
\cite{BerezinMarinov:1977,Casalbuoni:1976,BarducciEtal:1976,BrinkEtal:1976,%
BrinkEtal:1977,RietdijkHolten:1989,GibbonsEtal:1993,Tanimoto:1995,AhmedovAliev:2009,NgomeEtal:2010}. 
In $n$ number of spacetime dimensions, we
denote the particle's worldline coordinates by $x^\alpha$ ($\alpha=1,\dots , \dm$) and
describe its spin by a Lorentz vector of
Grassmann-odd coordinates  $\theta^a$ ($a = 1, \dots, \dm$), where $a$ is a vielbein index. 
That is, 
the motion of a particle is described by a curve $\tau \mapsto ( x(\tau), \theta(\tau)),  \tau \in \mathbb{R}$ and is governed by the following equations of
motion: 
\ba
\frac{D^2x^\alpha}{d\tau^2}&=&\ddot{x}^\alpha + \Gamma^\alpha_{\beta\gamma} \dot{x}^\beta \dot{x}^\gamma = \frac{i}{2} R^\alpha_{\ \beta ab}\theta^{a}\theta^b\dot x^\beta\,,
\label{eq:Papapetrou}\\
\frac{D \theta^a}{D\tau} &=& \dot{\theta}^a + \omega_\beta{}^a{}_b \dot{x}^\beta \theta^b = 0 \, . \label{eq:covariantly_constant_spin}
\ea
Here $\Gamma^\alpha_{\beta\gamma}$ and $\omega_\beta{}_{ab}$ are the Levi-Civita and spin connections, respectively, and 
$R_{\alpha\beta\gamma\delta}$ is the Riemann tensor.
The first equation is an analogue of the classical general-relativistic Papapetrou's equation, which generalizes the geodesic equation for an extended object with spin, whereas the latter 
equation expresses the simple requirement that,   
in absence of interactions other than gravity, the spin is constant along the motion of the particle. 
The equations of motion can be derived from the Lagrangian
\be
{\cal L}=\frac{1}{2}g_{\alpha\beta}\dot x^\alpha\dot x^\beta + \frac{i}{2}\eta_{ab}\theta^a\frac{D\theta^b}{d\tau}\,.
\ee

The theory possesses a generic {\em supercharge} $Q$,
\be\label{Qdef}
Q=\theta^a e_a{}^\alpha \Pi_\alpha\,,
\ee
which obeys
\be
\{H,Q\}=0\,, \quad \{Q,Q\}=-2iH\,.
\ee
Here, $H$ is the Hamiltonian,
\ba
H=\frac{1}{2}\Pi_\alpha\Pi_\beta g^{\alpha\beta}\,,\ \ 
\Pi_\alpha=p_\alpha - \frac{i}{2}\theta^a\theta^b\omega_{\alpha ab}=g_{\alpha\beta}\dot x^\beta\,,\qquad
\ea
$e_a^{\ \alpha}$ denotes the vielbein,
$p_\alpha$ is the momentum canonically conjugate to $x^\alpha$ and the Poisson brackets are defined as
\be\label{brackets}
\{F,G\}=\frac{\partial F}{\partial x^\alpha}\frac{\partial G}{\partial p_\alpha}-
\frac{\partial F}{\partial p_\alpha}\frac{\partial G}{\partial x^\alpha}+
i(-1)^{a_F}\frac{\partial F}{\partial \theta^a}\frac{\partial G}{\partial \theta_a}\,,
\ee
where $a_F$ is the Grassmann parity of $F$. For practical calculations it is useful to
rewrite these brackets in a covariant form \cite{GibbonsEtal:1993}
\ba\label{brackets2}
\{F,G\}&=&D_\alpha F\frac{\partial G}{\partial \Pi_\alpha}-
\frac{\partial F}{\partial \Pi_\alpha}D_\alpha G
+\frac{i}{2}\theta^a\theta^b R_{ab\alpha\beta}\frac{\partial F}{\partial \Pi_\alpha}\frac{\partial G}{\partial \Pi_\beta}\nonumber\\
&&+i(-1)^{a_F}\frac{\partial F}{\partial \theta^a}\frac{\partial G}{\partial \theta_a}\,,
\ea
where we have used the phase space covariant derivative
\ba
D_{\!\alpha} F=\partial_\alpha F - \omega_{\alpha ab}\theta^b\frac{\partial F}{\partial \theta_a}
+\Gamma_{\alpha\beta}^{\gamma}\Pi_\gamma\frac{\partial F}{\partial \Pi_\beta}\,\,.
\ea

Equations of motion are accompanied by two physical (gauge) conditions
\be\label{gaugecond}
2H=-1\,,\quad Q=0\,,
\ee
which state that $\tau$ is the proper time and the particle's spin is spacelike.

An important role for the spinning particle in curved spacetime is played by {\em non-generic superinvariants}, i.e., quantities 
that Poisson commute with the generic supercharge. 
More specifically, a superinvariant $S$ is defined by the equation
\be\label{Q}
\{Q, S\}=0\,.
\ee
Existence of solutions of such an equation imposes nontrivial conditions on the manifold. (The manifold has to possess special symmetries such as 
Killing vectors or Killing--Yano tensors,  for example.) It follows from the Jacobi identity that any superinvariant is automatically a constant of motion,  $\{H, S\}=0$. At the same time quantity $\{S,S\}$ is a `new' superinvariant and a constant of motion (which may, or may not be equal to $H$).
Hence, superinvariants 
correspond to an enhanced {\em worldline (super)symmetry}.

{\em Linear in momentum} superinvariants were studied in \cite{GibbonsEtal:1993, Tanimoto:1995}. 
In particular, when the following ansatz for the superinvariant is used:
\ba\label{ansatz}
Q_\omega&=&\theta^{a_1}\!\dots \theta^{a_{p-1}} \omega^\alpha{}_{a_1\dots a_{p-1}}\Pi_{\alpha} \nonumber\\
&&-\frac{i}{(p+1)^2}\theta^{a_1}\!\dots \theta^{a_{p+1}} \tilde \omega_{a_1\dots a_{p+1}}\,,
\ea
one finds that $p$-form $\omega$ has to satisfy the Killing--Yano equation 
\be\label{KY}
\nabla_\gamma \omega_{\alpha_1\dots \alpha_p}=\nabla_{[\gamma}\omega_{\alpha_1\dots \alpha_p]}=\frac{1}{p+1}(d\omega)_{\gamma \alpha_1\dots \alpha_p}\,, 
\ee
and that $\tilde \omega=d\omega$. Hence, to any Killing--Yano tensor $\omega$ there is a corresponding superinvariant $Q_\omega$ given by \eqref{ansatz}. Conversely, any superinvariant linear in momenta obeying \eqref{Q} 
is given by a Killing--Yano tensor and takes the form \eqref{ansatz} \cite{GibbonsEtal:1993, Tanimoto:1995}. 

In what follows we shall construct {\em quadratic in momenta} superinvariants and demonstrate that they enable us to integrate the bosonic sector of the 
spinning particle motion in the general rotating black hole spacetimes in higher dimensions described in the next section.


\section{Rotating black hole spacetimes}\label{sec3}
We are interested in the motion of a classical spinning particle in the vicinity of a general rotating Kerr-NUT-(A)dS black hole spacetime in higher dimensions 
\cite{ChenEtal:2006cqg}. We parametrize the total number of spacetime dimensions as $n=2N+\varepsilon$, where $\varepsilon=0,1$ in even and odd dimension, respectively. The euclidean version of the metric can be written in an orthonormal form
\begin{equation}\label{diagmetric}
{g}= \sum_{\mu=1}^N\,
    \Bigl(\,\ef^\mu \otimes \ef^\mu
    + \ef^{\hat \mu} \otimes \ef^{\hat\mu}\,\Bigr) + \eps \ef^0 \otimes \ef^0
  \;,
\end{equation}
where we have introduced the basis $\ef^a={\{\ef^\mu,\ef^{\hat\mu}, \ef^0\}}$,
\begin{equation}  \label{formframe}
\begin{gathered}
\ef^\mu = \frac{d x_{\mu}}{\sqrt{Q_\mu}}\;,\quad
  \ef^{\hat\mu} = \sqrt{Q_\mu} \sum_{j=0}^{N-1}\A{j}_{\mu}d\psi_j   \;, \\
E^0 = \sqrt{S} \sum_j \A{j}d\psi_j  \, .
\end{gathered}
\end{equation}
Here, coordinates $x_\mu\, (\mu=1,\dots,N)$\footnote{In the following Greek indices of the type $\mu, \nu, \dots$, from the middle of the alphabet, will take values $1, \dots , N$, while Greek indices of the type $\alpha, \beta$, from the beginning of the alphabet, will take values $1, \dots, \dm$.}  stand for the (Wick rotated) radial coordinate and longitudinal angles, and Killing
coordinates $\psi_k\; (k=0,\dots,N-1 +\eps)$ denote time and azimuthal angles associated with Killing vectors
${\KV{k}}$
\begin{equation}\label{KV}
\KV{k}=\cv{\psi_k}\;.
\end{equation}
We have further defined the functions
\ba
Q_\mu&=&\frac{X_\mu}{U_\mu}\,,\quad U_{\mu}=\prod\limits_{\nu\ne\mu} (x_{\nu}^2-x_{\mu}^2)  \;,\quad S = \frac{-c}{\A{N}} \, ,\label{eq:UandS_def}\\
\A{k}_{\mu}&=&\hspace{-5mm}\!\!
    \sum\limits_{\substack{\nu_1,\dots,\nu_k\\\nu_1<\dots<\nu_k,\;\nu_i\ne\mu}}\!\!\!\!\!\!\!\!\!\!
    x^2_{\nu_1}\cdots\, x^2_{\nu_k}\;,\ \
\A{j} = \hspace{-5mm} \sum\limits_{\substack{\nu_1,\dots,\nu_k\\\nu_1<\dots<\nu_k}}\!\!\!\!\!\!
    x^2_{\nu_1}\cdots\, x^2_{\nu_k}\; .\label{eq:A_def}\quad
\ea
The quantities ${X_\mu}$ are functions of a single variable ${x_\mu}$, and $c$ is an arbitrary constant.
The vacuum (with a cosmological constant) black hole geometry is recovered by setting
\begin{equation}\label{BHXs}
  X_\mu = \sum_{k=\eps}^{N}\, c_{k}\, x_\mu^{2k} - 2 b_\mu\, x_\mu^{1-\eps} + \frac{\eps c}{x_\mu^2} \; .
\end{equation}
The constant $c_N$ is proportional to the cosmological constant and the remaining constants are related to angular momenta,
mass, and NUT parameters.

The dual vector frame $\ef_a={\{\ef_\mu,\ef_{\hat\mu}, \ef_0\}}$ reads
\begin{gather}\label{Es}
 \ef_\mu = \sqrt{Q_\mu}\cv{x_{\mu}}\;,\quad
 \ef_{\hat\mu} = \sqrt{Q_\mu}\sum_{j}\frac{(-x_\mu^2)^{N{-}1{-}j}}{X_\mu}\cv{\psi_j} \; , \nn\\
 E_0 = \frac{1}{\sqrt{S} \A{N}} \cv{\psi_N} \, . \label{vectfr}
\end{gather}

Besides the Killing vectors $\KV{k}$, \eqref{KV}, the metric \eqref{diagmetric} possesses a hidden symmetry of the principal conformal Killing--Yano tensor \cite{FrolovKubiznak:2007, KubiznakFrolov:2007, KrtousEtal:2007jhep}. In the basis \eqref{formframe} the principal conformal Killing--Yano 2-form reads
\begin{equation}\label{PCKY}
    h = \sum_\mu x^\mu\, \ef^\mu\wedge\ef^{\hat\mu}\;.
\end{equation}
This tensor generates the tower of Killing--Yano tensors 
[obeying \eqref{KY}]
\be\label{fj}
f_{(j)}=*h^{\wedge j}\equiv\frac{1}{j!\sqrt{(n-2j-1)!}}*\bigl( \underbrace{{h}\wedge \ldots \wedge
{h}}_{\mbox{\tiny{$j$-times}}}\bigr)\,,
\ee
which, in their turn, `square' to the
2nd-rank Killing tensors
\begin{equation}\label{KT}
\KT{j}=\sum_{\mu}\A{j}_\mu
\Bigl(\ef^\mu \otimes  \ef^\mu\! +\!
\ef^{\hat \mu}\otimes\ef^{\hat \mu}\Bigr) + \eps \A{j} \ef^0 \otimes  \ef^0   \;.
\end{equation}
Obviously, $\KT{0}$ coincides with the metric and hence it is a trivial Killing tensor which we include in our tower; so we take $j=0,\dots, N-1$.

It was demonstrated in \cite{PageEtal:2007, FrolovEtal:2007} that explicit symmetries $\KV{k}$ and hidden symmetries $\KT{j}$ guarantee complete integrability of geodesic motion  
in these spacetimes. We shall now demonstrate that this result can be extended to the spinning particle theory.

\section{Quadratic superinvariants}
The rotating black hole spacetime \eqref{diagmetric} admits naturally the following $(N+\varepsilon)$  
bosonic linear in velocities superinvariants \eqref{ansatz} corresponding to the isometries $\xi_{(k)}$, \eqref{KV}:
\be\label{Qk}
Q_{\xi_{(k)}}=\xi_{(k)}^\alpha\Pi_\alpha-\frac{i}{4}\theta^a\theta^b (d\xi_{(k)})_{ab}\,.
\ee 

Besides these it also admits $N$ linear in velocities superinvariants  \eqref{ansatz} associated with the 
Killing--Yano symmetries \eqref{fj}. However, such superinvariants are not what we are after because of two reasons.
First, such invariants are not in involution. In fact one can show that in even dimensions, where such superinvariants are fermionic, their Poisson brackets 
do not close and generate  an extended superalgebra, c.f. \cite{AhmedovAliev:2009}. Second, such integrals are not 
`invertible' for velocities. This stems from the defining property of the Grasmann algebra and the fact that 
the term in \eqref{ansatz} involving velocities contains also $\theta$'s. 

To circumvent these difficulties we seek new bosonic superinvariants ${\cal K}_{(j)}$ such that their leading (quadratic in velocities) term contains no $\theta$'s and is completely determined by the Killing tensors $K_{(j)}$, \eqref{KT}:
\ba\label{quadr}
{\cal K}_{(j)}&=&K_{(j)}^{\alpha\beta} \Pi_\alpha\Pi_\beta+{\cal L}_{(j)}^\alpha\Pi_\alpha+{\cal M}_{(j)}\,,\nonumber\\
{\cal L}^\mu_{(j)}&=&\theta^{a}\theta^b L_{(j) ab}{}^\mu\,,\quad  
{\cal M}_{(j)}=\theta^{a}\theta^b\theta^c\theta^d M_{(j)abcd}\,.\quad
\ea
In the absence of spin, such quantities reduce to the quadratic integrals of geodesic motion, responsible for its complete integrability. 

A quantity ${\cal K}$ of the form \eqref{quadr} is a superinvariant (for the moment we have dropped the index $j$), i.e., obeying 
\be
\{Q, {\cal K}\}=-\theta^\alpha D_\alpha {\, \cal K}-i\Pi^\alpha\frac{\partial {\cal K}}{\partial \theta^\alpha}=0 \, , 
\ee
provided that the following three equations are satisfied:
\ba\label{eqQ}
\nabla_\gamma K_{\alpha\beta}+2iL_{(\alpha|\gamma|\beta)}=0\,,\nonumber\\
4iM_{\gamma\delta \alpha\beta}+\nabla_{[\delta}L_{\alpha\beta]\gamma}=0\,,\\
dM=0\,.\nonumber
\ea
Obviously, the last condition is a trivial consequence of the second and we do not have to discuss it anymore.
The second equation implies that $4iM_{\gamma\delta \alpha \beta}=\nabla_{[\gamma}L_{\delta \alpha\beta]}$, i.e., 
once $L_{\alpha\beta\gamma}$ is known then $M_{\alpha\beta\gamma\delta}$ is also determined.
The remaining information in the second equation expresses the requirement that  
\be\label{nontrivial}
\nabla_{[\delta}L_{\alpha\beta]\gamma}=\nabla_{[\delta}L_{\alpha\beta\gamma]}\,.
\ee
Since $L_{\alpha\beta \gamma}$ is not completely antisymmetric, such a condition is highly non-trivial. 

In the Kerr-NUT-(A)dS spacetimes we try the following ansatz for a solution of these equations:
\ba\label{solution}
K^{\alpha\beta}&=&f^{\alpha \kappa_1\dots \kappa_{p-1}}f^\beta{}_{\kappa_1\dots \kappa_{p-1}}\,,\label{solK}\nonumber\\
L_{\alpha\beta}{}^\gamma&=&A f_{[\alpha|\kappa_1\dots \kappa_{p-1}|}(df)_{\beta]}{}^{\gamma\kappa_1\dots \kappa_{p-1}}\nonumber\\
&+&\vspace{0.5cm}B(df)_{\alpha\beta \kappa_1\dots \kappa_{p-1}}f^{\gamma\kappa_1\dots \kappa_{p-1}}\,,\qquad \label{solL}\\
M_{\alpha\beta\gamma\delta}&=&-\frac{i}{4} \nabla_{[\alpha} L_{\beta\gamma\delta]}\,.\nonumber\label{solM}
\ea
Here, $f_{\alpha_1\dots \alpha_p}$ is a corresponding rank-$p$ Killing--Yano tensor and constants $A$ and $B$ are coefficients to be determined.\footnote{%
This formula generalizes the formula given in \cite{GibbonsEtal:1993} for a $p=2$ Killing--Yano tensor $f$, in which case 
one can obtain the expression for ${\cal K}$ as 
\be
{\cal K}=\{Q_f,Q_f\}
\ee
 with $Q_f$ being the linear superinvariant \eqref{ansatz}, and one has $A=B=-2i/3$.
For higher-rank Killing--Yano tensors $f$ we are not aware of any corresponding relation. 
}

Using the explicit form of the Kerr-NUT-(A)dS metric \eqref{diagmetric} and Killing--Yano tensors $f_{(j)}$, \eqref{fj}, therein
one can show \cite{CarigliaKubiznak:2011} that 
Eq. \eqref{nontrivial} is satisfied
when $A=B$. The first Eq. \eqref{eqQ} then fixes the total coefficient and we recover 
\be
A=B=-\frac{2i}{p+1}\,.
\ee
With this choice of coefficients the quantities ${\cal K}_{(j)}$, given by \eqref{quadr} and \eqref{solution}, are superinvariants and hence also constants of spinning particle motion 
in Kerr-NUT-(A)dS spacetimes in any dimension.\footnote{It is interesting to remark here that in order to prove this fact it is not enough to consider the Killing--Yano 
equation and its integrability condition; one has to use the explicit form of the metric as well as of the Killing--Yano tensors therein.}

\section{Integrability}
We shall now use the newly constructed superinvariants to prove Liouville integrability of the bosonic equations of the spinning particle motion.  Our conjecture is that in Kerr-NUT-(A)dS spacetimes in any dimension the following quantities:
\be\label{integrals}
\{H, {\cal K}_{(1)},\dots ,{\cal K}_{(N-1)}, Q_{\xi^{(0)}},\dots ,Q_{\xi^{{(N-1+\eps)}}}\}\,,
\ee
form a complete set of bosonic integrals of spinning particle motion that are 
functionally independent and in involution with respect to the Dirac-Poisson brackets \eqref{brackets}---making this 
motion completely integrable.

The functional independence of these constants of motion follows from the independence of the corresponding Killing vectors $\xi_{(k)}$, \eqref{KV}, and Killing tensors $K_{(j)}$, \eqref{KT}; c.f. \cite{PageEtal:2007}.

 To prove the involution we have to show that for any $i, j, k, l$ one has  
\be\label{BR}
\{Q_{\xi_{(k)}}, Q_{\xi_{(l)}}\}=0\,,\ 
\{Q_{\xi_{(k)}}, {\cal K}_{(i)}\}=0\,,\ 
\{{\cal K}_{(i)}, {\cal K}_{(j)}\}=0\,.
\ee
Each bracket corresponds to several terms, characterized
by a number of fermionic and momentum coordinates, which have to vanish separately.
In the zeroth-order in $\theta$'s one recovers the {\em Schouten--Nijenhuis brackets} of the corresponding tensors. 
All such brackets vanish due to the result on complete integrability of geodesic motion.
The first Poisson bracket \eqref{BR}, $\{Q_{\xi_{(k)}}, Q_{\xi_{(l)}}\}$, contains an additional term quadratic in $\theta$'s. Using the integrability condition for Killing vectors and various Bianchi identities  one can show
that this term automatically vanishes when the two Killing vectors commute, i.e., when the zeroth-order term vanishes. Hence we have shown that the first condition \eqref{BR} is satisfied in all dimensions. 

The situation is  more complicated with the other two brackets in \eqref{BR}. The second condition in \eqref{BR} imposes two highly non-trivial equations, the third imposes  three equations (see \cite{CarigliaKubiznak:2011} for more details). One can show that such equations are not automatically satisfied when the corresponding Schouten--Nijenhuis brackets vanish. Hence they represent 
additional requirements (in addition to the requirements imposed by complete integrability of geodesic motion) which have to be satisfied in order for the bosonic part of spinning particle motion to be integrable. Due to their complexity we have not tried to verify such equations analytically. However, we have checked that they are satisfied in 4, 5, 6, and 7 dimensions using the Maple software, proving complete integrability in these cases. Since there are no special features that depend on the actual dimension of spacetime we have reasons to strongly believe that all three equations \eqref{BR} are valid in any dimension.

Let us finally write down the expressions for the spinning particle momenta in the tetrad basis.
Writing  
\be
p=p_\mu\ef^\mu+p_{\hat \mu}\ef^{\hat \mu}+\varepsilon p_{ 0} \ef^{ 0}\,,
\ee
we find that the momenta can be written in terms of the constants of motion $\Psi_k\equiv Q_{\xi_{(k)}}$, $\kappa_j\equiv {\cal K}_{(j)}$, $\kappa_0\equiv H=-1/2$, as follows, c.f., \cite{KrtousEtal:2007jhep}:
\ba 
p_{\hat \mu}&=&\sqrt{Q_\mu}\sum_k\frac{(-x_\mu^2)^{N-1-k}}{X_\mu} \Bigl(\Psi_k+\frac{i}{4}\theta^{a}\theta^b(d\xi_{(k)})_{ab}\Bigr)\,,\nonumber\\
p_{0}&=&\frac{1}{\sqrt{S}A^{(N)}}\Bigl(\Psi_N+\frac{i}{4}\theta^{a}\theta^b(d\xi_{(N)})_{ab}\Bigr)\,.
\ea
The last components of the momentum, $p_{\hat \mu}$, are given by the solution of the (unfortunately coupled) system of equations
\ba
\kappa_j&=&\sum_\mu\Bigl[ A_\mu^{(j)}(p_\mu^2+p_{\hat \mu}^2)+{\cal L}^\mu_{(j)}p_\mu+
{\cal L}^{\hat \mu}_{(j)}p_{\hat \mu}\Bigr]\nonumber\\
&&+\varepsilon \Bigl(A^{(j)}p_{0}^2+{\cal L}^{0}_{(j)}p_{0}\Bigr)+{\cal M}_{(j)}\,,
\ea
with ${\cal L}^\alpha_{(j)}$ and ${\cal M}_{(j)}$ given by \eqref{quadr} and \eqref{solution}.

\section{Discussion}
In this paper we have constructed new bosonic quadratic in velocities integrals of spinning particle motion in higher-dimensional rotating black hole spacetimes of any dimension. We have further demonstrated that in 4, 5, 6, and 7 dimensions such integrals together with the integrals corresponding to Killing symmetries are functionally independent one of another and mutually Poisson commute---hence they guarantee Liouville  integrability of the bosonic part of the spinning particle motion in this case. We conjecture that this remains true in any dimension. Let us emphasize that in this paper we have not dealt with the fermionic part of the motion, its integrability would require a separate analysis. Hence the question of complete integrability of the whole (bosonic and fermionic) system of spinning particle equations of motion remains open. We might comment, however, that there are reasons to suspect that the system might be fully integrable. The Dirac equation in fact, which corresponds to the quantised system, is solvable in these metrics and admits separation of variables \cite{OotaYasui:2008,CarigliaEtal:2011_04}. Also to achieve such separation of variables it is enough to use a set of $\dm$ mutually-commuting operators, as many as the Poisson commuting functions that we have discussed in this paper.

The obtained results raise various interesting questions. Is it possible to formulate the Hamilton--Jacobi theory for the spinning particle and derive the demonstrated integrability as a consequence of separability of the Hamilton--Jacobi equation, similar to the geodesic case \cite{Carter:1968pr, FrolovEtal:2007}? 
Even more interestingly, is it possible to exploit the close relationship between the discussed spinning particle theory and the classical general-relativistic theory of spinning objects described by Papapetrou's equations? Formally, such equations are obtained by replacing $- i \theta^a\theta^b$ with the spin tensor $S^{ab}$. [It can be explicitly checked that $-i \theta^a\theta^b$ satisfy the correct Lie algebra of the Lorentz group under Poisson brackets.]
After this identification Eqs. \eqref{eq:Papapetrou} and \eqref{eq:covariantly_constant_spin} become Papapetrou's equations
with the particular choice of supplementary condition:
\be\label{Pap}
\frac{D^2x^\mu}{d\tau^2}= - \frac{1}{2} R^\mu_{\ \nu\kappa\lambda}S^{\kappa\lambda}\dot x^\nu\,,\quad \frac{DS^{\mu\nu}}{d\tau}=0\,.
\ee 
Under such a transition, linear superinvariants \eqref{Qk} translate into the full integrals of motion for Papapetrou's equations \eqref{Pap}. However, 
even in the Kerr-NUT-AdS black hole spacetimes the new quadratic superinvariants \eqref{quadr} become only approximate integrals---to a linear order in the spin tensor $S^{ab}$. 
This highlights 
the difference between the two theories. An interesting question is: is it possible to `upgrade' these invariants to provide full quadratic integrals of motion in Papapetrou's theory? 
Similarly, can we infer some properties of the Dirac equation in these spacetimes? 
Dirac's theory is formally recovered when one replaces $\theta$'s with $\gamma$ matrices and $\Pi$'s with the spinorial derivative. A natural question is: are the second order operators corresponding to superinvariants \eqref{quadr} symmetry operators of the Dirac operator?  These are just a few interesting questions left for future work.

\vspace{0.2cm}

\section*{Acknowledgments}
We are grateful to V.P. Frolov, G.W. Gibbons, P. Krtou\v{s}, D. Sorokin, and C.M. Warnick for useful discussions and reading the manuscript.
D.K. acknowledges the Herchel Smith Postdoctoral Fellowship at the University of Cambridge. M.C. acknowledges partial financial support from Funda\c c\~ao de Amparo \`a Pesquisa de Minas Gerais - FAPEMIG. 



\providecommand{\href}[2]{#2}\begingroup\raggedright\endgroup

\end{document}